# Assisted Requirements Selection by Clustering

José del Sagrado, Isabel M. del Águila

Dpt. of informatics,

University of of Almer´ıa, Spain.

jsagrado@ual.es

**Abstract**

Requirements selection is a decision-making process that enables project managers to focus on the deliverables that add most value to the project outcome. This task is performed to define which features or requirements will be developed in the next release. It is a complex multi-criteria decision process that has been focused by many research works because a balance between business profits and investment is needed. The spectrum of prioritization techniques spans from simple and qualitative to elaborated analytic prioritization approaches that fall into the category of optimization algorithms. This work studies the combination of the qualitative MoSCoW method and cluster analysis for requirements selection. The feasibility of our methodology has been tested on three case studies (with 20, 50 and 100 requirements). In each of them, the requirements have been clustered, then the clustering configurations found have been evaluated using internal validation measures for the compactness, connectivity and separability of the clusters. The experimental results show the validity of clustering strategies for the identification of the core set of requirements for the software product, being the number of categories proposed by MoSCoW a good starting point in requirements prioritization and negotiation.

## 1 Introduction

Requirements engineering identifies, documents, negotiates and manages the desired features and constraints of software-intensive systems and the assumptions about the environment [8]. The most serious problems in complex systems development frequently arise from requirements engineering [19, 16]. The processes related to requirements are knowledge intensive tasks that are often supported by human effort and expertise. Requirements are the glue that keeps together the stages in project development because they collect the needs or conditions that the product under construction should meet to succeed.

A requirements engineering decision process, in software product development, is the definition of which features or requirements will be developed in the next release under technical, resource, risk, and budget constraints [17]. It is a complex multi-criteria decision process that, most times, entails achieving a balance between the value the requirements add to the project outcome and their cost. A clear instance are agile projects, because requirements and solutions quickly evolve through the collaborative teams and customers effort, making requirements selection a critical process.

Requirements selection has been the focus of attention in many research works, and the software industry [7, 40, 1, 41, 48, 23]. The proposed techniques use different information associated to requirements. Some proposals use only one attribute, others a combination of them. Besides, both stakeholders' or developers' side could be considered inside the process. The range of suitable techniques to be applied is broad and diverse. However, both a deep knowledge of the domain and a skilful quantification capability have to be involved to succeed at using most of these techniques [8].

Although requirements engineering decision processes have been typically based both on the stakeholders' intuition and experience and on rational schemes (such as criteria, options or arguments), the recent soaring of data driven approaches in requirement engineering [30] puts pressure on practitioners to use and integrate quantitative data to decide automatically on what requirements and features should be added to or removed from future releases. The use of experiments, case studies, surveys, and whatever available quantitative data is being extrapolated from software engineering empirical approaches, which emphasize the use of all kinds of empirical studies to accumulate knowledge in requirements related tasks. [48]

We investigate if some kind of synergy can be found between two well-known methods, MoSCoW and clustering, as a model to automatically support the problem of requirements selection. MoSCoW is a qualitative technique for requirements prioritization which is based on the classification of requirements using plain English meaning of the prioritization categories. It helps stakeholders to make up their minds about the next release objectives and it has the value of providing them a better understanding of the impact of their categorization decision. Clustering methods aim to partition $n$ observations into $k$ clusters in such a way that observations in one cluster are more similar to each other than those in other clusters. The notion of similarity used by clustering algorithms is related to that of distance and is based on quantitative data. There are several measures associated to requirements, such as development effort and stakeholders' satisfaction, which can be used for finding out requirements groups.

The research has been conducted stating a goal, which brings us into some research questions and, finally, carrying out some empirical experiments, we get the information needed to give the answers to the research questions. Our goal is to check if available quantitative records about requirements, in particular estimated development effort and clients' satisfaction are reliable enough to define automatically the core set of requirements in a software development project (i.e. the mandatory requirements for a successful software product).

Based on this goal, we derived three research questions:

- **RQ1** Can quantitative data about requirements (such as satisfaction and effort) be used to cluster and prioritize requirements automatically?
- **RQ2** Which clustering algorithm gets better results for requirements selection?
- **RQ3** Can the obtained requirements clusters be interpreted to assist in the selection of the requirements set to be developed?

The structure of the rest of the paper starts by describing the background of well-known techniques for requirements selection in Section 2. Then, Section 3 describes the quantitative formulation of the selection problem using development effort and customers' satisfaction and the two methods to be combined, MoSCoW and clustering. Section 4 is devoted to defining the process followed to check the suitability of the combined approach and how to set the number of clusters and map MoSCoW categories to clusters. The results of the effectiveness study of the proposal for three different data sets is shown and discussed in Section 5. Section 6 addresses the limitations and threats to validity. Finally, Section 7 includes the conclusions.

## 2 Background

Software project planning strategies are becoming the cornerstones of software industries because they not only define the features to be built but also exactly when they will be released and help to manage cost. The Standish Group annually reported that around 80% of surveyed software projects do not achieve their definition of success based on time, cost and scope criteria [16]. In fact, since 2015 the scope criterion, which represents how requirements have been fulfilled, has been relaxed. They also reported that one of the most relevant reasons for software project to fail, is shifting requirements, because requirements are often documented textually and the requirements specifications themselves are rarely changed [45]. This suggests that software projects fail is due to their inability to evolve efficiently to match the shifting requirements or to afford the new ones that appear in project evolution, reinforcing the importance of releases management. In consequence, making a proper decision about what functionality a release of an evolving software product should have, is critical for the success or failure of the whole project. The better the selection made is, the fewer problems with shifting requirements would be found in next releases.

To answer what to release, many approaches represent the problem as a constraint problem or as an analytic study of the features values, features dependencies, or even stakeholder' priorities or dislikes. Researchers in requirements triage, requirements prioritization, requirements selection, next release problem or release planning have been prolific from their origin [26, 5, 13, 48, 14]. Each one, directly or indirectly, contributes to model different points of view and devises diversified solving methods.

When a software product is being developed, especially if agile methodologies are applied, requirements prioritization and selection are recurrent activities. Not only is it a process to identify and filter the important requirements, but also to solve conflicts and plan the different product deliveries. These complex decisions require a detailed knowledge of the domain and good quantification and estimation techniques of the requirements properties, also involving contradictory criteria [9]. However, the variety of prioritization methods makes it hard to select the most useful one.

Various factors and dimensions can be considered for requirements selection. Some of them are defined either by customers or stakeholders (e.g. requirement perceived value, deadline), others by the development team (e.g. available effort, team size), or maybe by both (e.g. risk, volatility) or, going further, by external factors such as market value issues.

Many ranking techniques and prioritization techniques have been defined, each one using a subset of the information collected for requirements [41, 23, 10] or applying only people expertise. These methods may differ in the way priorities are computed, in the scale of values used to represent the resulting ordering and in the accuracy of the results. Classifying them is difficult because several classification criteria could be used, either conjointly or independently.

Some methods are ad-hoc, others use the attributes, mainly quantitative, that characterize requirements. There are techniques that manage only one attribute, but others use a combination of them, considering dependencies between requirements or not. Other techniques study only the client's position, only the team point of view, or both. Even their goals can be sightly different, since some methods select which are the requirements that should be included or rejected, or how to distribute them in releases. Prioritization approaches list is wide, even more because some of them can be combined. Next the most representative are briefly revised.

One well-known technique that does not use quantitative data about requirements attributes is the MoSCoW method. It proposes arranging requirements in four categories based on expert judgment and the elicited information. The term MoSCoW is derived from the first letter of each of four categories (Must have, Should have, Could have and Won't have), embedding a semantic meaning, as well [9]. Top-Ten technique can be used when a completely sorted or prioritized list of requirements is needed. The 10 most important requirements are selected by the clients.

The Kano model provides a different method to assist developers to understand customers' perspectives on product features by assessing two measures for each candidate requirement: the satisfaction and dissatisfaction. The responses to these two measures will arrange requirements into different scoring categories [25].

Another type of methods are those that use a numerical approach but without mapping numbers to specific quantitative attributes. Simple ranking is one of them. The requirements have to be sorted based on the prioritization criterion. For quantitative or ranked criteria this approach becomes a sort problem, but the final ranking has to reach an agreement among how each client or ex-pert ranks requirements. Cumulative voting method allows voters to distribute an explicit number of points amongst requirements. In the 100-point method, 100 units are given to stakeholders to cast their vote for their most valuable requirements [28]. Planning games approaches [6] actively involves stakeholders in ranking processes. Time-boxing technique assess the amount of work that the project team can deliver during the prescribed period. It is a way of focusing on achieving what needs to be done without delay or procrastination based on the effort bound, by selecting the requirements that fit in this bound.

More formal and systematic approaches are the analytic hierarchy process (AHP) and prioritization matrix [46] that use more than two criteria to prioritize requirements. AHP evaluates requirements using only one attribute (e.g. effort, value, risk) by doing a pairwise comparison in a square matrix. It converts these values to a total order relationship between the requirements for this attribute [26]. This technique can be extended to combine more than one attribute or criterion. Another cost-value method combines AHP and a graphical approach for two attributes, value and effort. It has as the overall goal to select requirements that give maximum value for minimum cost within allowable cost limits of a software system, by applying a variant of 0-1 knapsack model [24].

Requirements selection problem has been also defined as an optimization problem, called next release problem [5], that lately has been reformulated as a multi-objective problem where contradictory objectives can be defined. Search Based Software Engineering (SBSE) combines computational methods, that use the estimated attributes of a software artefact, with expert human expertise to achieve best ranking results [3, 20, 48], giving automatic support to requirements selection. Not only release related problems but also many processes in software engineering can be formulated as optimization problems throughout the Software Engineering life cycle, such as non-dominated sorting of the component to be reused, or the process of automatically generating test data according to a test adequacy criterion using search-based optimization algorithms. These approaches apply meta-heuristic search techniques from requirements and project planning to maintenance and re-engineering, [21].

Most SBSE works that deal with requirements, focus on the problem that involves the determination of what features or requirements should be covered by the software product under construction based on the cost-value criterion [24]. The starting point is the set of candidates requirements (or product backlog) with their associated development effort and a defined number of clients, each one with their particular demands about what to include in the next version of the product under development.

## 3 Quantitative MoSCoW requirements selection

The strong semantic approach of MoSCoW method can be supported by the data (value, effort) postulated for the cost-value criterion[24], which is described in this section by combining the pioneering next release models [5, 44]. This quantitative formulation offers a suitable solution for the problem by applying clustering approaches, nonetheless the robustness of clustering algorithms could allow extending this formulation by using additional scoring dimensions for the problem, such as risks or volatility. These methods have a solid base with many successfully developed algorithms that can manage and prove the suitability of the intertwining, and the research questions proposed.

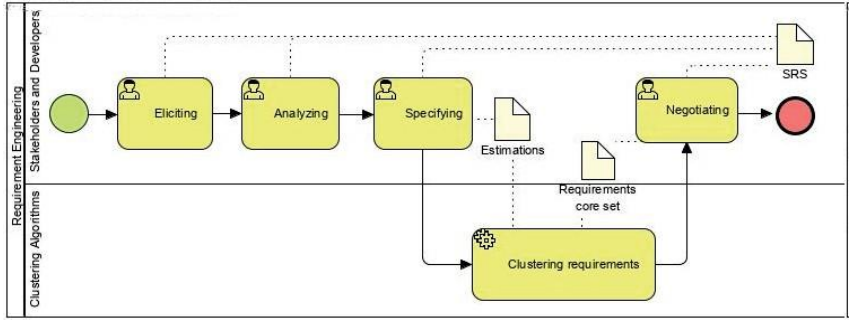

Figure 1: Workflow for requirements selection

Our purpose is to use clustering as an assistant in the critical decision process about which features or requirements will be developed in the next release. This task falls into the conflict resolution subtopic [8]. In most cases, it is unwise for the developers to make a unilateral decision, so it becomes necessary to make an agreement. It is often important, for contractual reasons, that such decisions be traceable back to the stakeholders. Figure 1 shows the workflow we propose to use clustering in this decision-making process. In the usual workflow, once the requirements are elicited and scored by stakeholders and developers, specification and analysis tasks should be executed. Analysis is where requirements prioritization is usually performed.

Our proposal extracts an automated clustering activity being in charge of the definition of the core set of requirements, those that belong to the *'must have'* category. According to the resources which are available, also called team capacity or effort bound, the next release objective could be easily negotiated starting from the obtained core set of requirements.

### 3.1 Quantitative problem formulation

Let $\mathbf{R} = \{r_1, r_2, \ldots, r_n\}$ be the set of requirements to be considered. These requirements represent enhancements or new features that are suggested by $m$ customers and are also candidates to be solved in the next software release. Customers are not equally important. So, each customer $i$ will have a weight $w_i$, which measures its importance for the software project. Let $\mathbf{W} = \{w_1, w_2, \ldots, w_m\}$ be the set of customers' weights.

Each requirement $r_j$ in $\mathbf{R}$ has an associated development effort $e_j$, which represents estimated cost for its development. Let $\mathbf{E} = \{e_1, e_2, \ldots, e_n\}$ be the set of requirements efforts. The same requirement can be suggested by several customers showing a different priority for each customer. Thus, the importance that a requirement $r_j$ has for customer $i$ is given by a value $v_{ij}$. The higher the $v_{ij}$ value, the higher is the priority of the requirement $r_j$ for customer $i$.

The added value given by the requirement inclusion $r_j$ in the next software release, also called global satisfaction, $s_j$, is measured as a weighted sum of its importance values for all the customers ($s_j = \sum_{i=1}^{m} w_i \cdot v_{ij}$). The set of requirements satisfactions is denoted as $\mathbf{S} = \{s_1, s_2, \ldots, s_n\}$.

Some extended models [34] also include dislikes about a requirement to take into account user opinion regardless of the effort that its development involve. We would rather maintain the widely agreed formulation of the next release problem [5, 44] putting in the same level the stakeholders' and developers' point of view.

The problem presents several constraints that should be fulfilled in the solution found. The one is due to the effort bound, also called cost limit or capacity $B$, which represents the amount of available development resources. Besides, requirements present different types of interactions or dependencies, which are also problem constraints. These interactions must be considered forcing us to check whether any conflict is present whenever we try to select a new requirement. This fact forces to implement the requirements in a specific order [26, 12]. Dependencies can be broken down into two groups. The first can be explicitly represented as a graph $G$ [35, 38] and comprises the functional dependencies: *Implication or precedence* ($r_i \Rightarrow r_j$) dependency stands for that the requirement $r_j$ cannot be selected if a requirement $r_i$ has not been implemented yet, *Combination or coupling* ($r_i \odot r_j$) represents that a requirement $r_i$ cannot be included separately from a requirement $r_j$, *Exclusion* dependency ($r_i \oplus r_j$) means that the requirement $r_i$ can not be included together with the requirement $r_j$. The second group (denoted as $\Delta S$ and $\Delta E$) includes those dependencies that imply changes in the amount of resources needed or the benefit related to each requirement: *Revenue-based* appears whether the development of a requirement $r_i$ implies that some others requirements will increase or decrease their value, *Cost-based* relationship carries that the development of a requirement $r_i$ implies some others requirements will increase or decrease their implementation cost.

These dependencies can be modelled as a pair of $n \times n$ symmetric matrices: $\Delta S$, where each element represents the increment or decrease of $s_j$ and $\Delta E$, that represents a change in the effort when implementation of two requirements take place in the same release [38]. So we have a 5-fold problem ($R, S, E, L, \hat{R}$), being $\hat{R}$ the set of requirements to be developed in the next release, that is the solution to the problem and $L = (B, G, \Delta S, \Delta E)$ the set of constraints to be considered.

## 3.2 MoSCoW method

Using a consistent language makes it easier to identify different classes of requirements. A simple example of this is the use of 'shall' as a keyword could indicate that a requirement is present in the written-down sentence. Some approaches go further and use 'shall,' 'should' and 'may' to indicate different priorities of a requirement. MoSCoW method reproduces this use of the language. Its basis is the fact that, even all requirements are important, the identification of vital requirements is a fundamental task. These requirements are those that contribute the most to the project success, and they are mandatory. If some of them are not accomplished, the final product won't be a viable product. Besides, including the more important functionalities first facilitates the reception of an incremental feedback from users, schedule adjustments, and the removal of errors and misunderstandings between developers and clients in early stages which leads to customer satisfaction.

Table 1: MoSCoW categories.

| Label | Meaning |
|---|---|
| **M** Must have | Contains the requirements that must be satisfied in the final solution for the solution to be considered a success. Must can be also thought of as a Minimum Usable SubseT |
| **S** Should have | Represents the high-priority items that should be included in the solution if it is possible. These are often critical requirements but which can be satisfied in other ways if strictly necessary. Should requirements are as important as Must but may not be time _critical or may have a work around |
| **C** Could have | Describes the requirements considered as desirable but unnecessary. These will be included if time and resources permit. |
| **W** Won't have | Represents the requirements that stakeholders have agreed will not be implemented in a release, but may be considered for the future. |

MoSCoW gets its name from an acronym formed by the following priority labels: Must have, Should have, Could have, and Won't have. The letter *o* is added to make the acronym pronounceable. The classifications are shown in table 1. MUST can also be considered an acronym for the Minimum Usable SubseT, because this category represents the compulsory requirements for the release, that is, those that have to be included in the software product in order to get clients' acceptance.

The main difference from other techniques, which classify using 'high', 'medium', or 'low', is that MoSCoW provides a semantic meaning. The stakeholders, in charge of the classification, know the real effect that their requirements assessment will produce. Human expertise and involvement is the basis of this qualitative prioritization method. MoSCoW provides a way to reach a common understanding on the relative importance of including a specific feature in the product. The categories define an agreed 'red line' that cannot be passed.

The pursued goal is to identify the set of requirements that will make up the core of a software product. That is to say, the set of fundamental requirements that the software to be developed 'must have' using as little human effort as possible. Could we match quantitative requirement attributes, such as effort or satisfaction, with MoSCoW categories to arrange requirements automatically or semi-automatically becoming the starting point for requirements negotiation? What we need next, is a protocol to make groups of requirements using their associated quantitative properties.

Some other well-known models also classify the properties or features of a product into categories, such as the Kano model. This method collects a set of ideas and techniques that help us determine customers' satisfaction with product features, by describing the connection between customer satisfaction and the realization of customer requirements. While many works establish three types of attributes to products and services (must be, one-dimensional, attractive) [39] or (dissatisfiers, satisfiers, delighters) [36], other works also manage the customers' emotions and satisfaction/dissatisfaction defining five or even six categories [32]. In consequence, the starting point will be the number of categories indicated by MoSCOW, but without taking it for granted.

### 3.3 Clustering

Nowadays, large amounts of data are being collected continuously, and it is becoming increasingly important to discover knowledge in multidimensional data. Here is where clustering methods come into play, because their goal is to discover groups of similar objects within a data set. Thus, clustering approaches require some methods for measuring the distance or the (dis)similarity between objects, so that objects in the same group should be closer to each other and further than those in other groups of objects. Common clustering approaches can be classified into one of the next bigger groups:

- *Partitioning methods* that divide the data into a pre-specified number of groups. *K-means* and *K-medoids* (i.e. partitioning around medoids algorithm) belong to this group. [31, 22]

- *Hierarchical methods* [33] that do not require pre-specifying the number of clusters to be generated. They return a dendogram, a tree-based representation of objects and groups. Here, *agglomerative clustering* is the most representative approach.

*K-means* algorithm [31, 22] is perhaps the most popular strategy to finding clusters in data. Each cluster is represented by its *centroid*, which is the mean of the observations assigned to the cluster. Initially, once the number $k$ of groups has been specified, $k$ observations are chosen as centroids for the clusters. The clusters are defined so that the distance between each observation in a cluster and its centroid is a minimum. Even though, *K-means* is simple, fast and can deal with large datasets, the number $k$ of clusters has to be specified in advance and the results obtained depend on the initial centroids selection. Besides, it is sensitive to the data ordering (i.e. if you rearrange your data, possibly you'll get a different solution) and, finally, the *K-means* method is also sensitive to anomalous data points and outliers.

Partitioning Around Medoids (PAM) [27] rests on the concept of *medoid*, which is an observation within a cluster such as the sum of the distances between it and all the other observations in the cluster is a minimum. Clusters are constructed by assigning each observation to the nearest medoid. Then, *PAM* tries to improve the quality of the clustering by interchanging medoids with the other observations and checking if the distances respect to the medoid are reduced. Although being more reliable and less sensitive to outliers, *PAM* requires more computation effort than *K-means*.

In contrast to partitioning methods (i.e. *K-means* and PAM), *agglomerative hierarchical clustering* groups observations based on their similarity and does not require pre-specifying the number of clusters to be produced. Initially, each observation is considered as a separate cluster (i.e. a leaf in the dendogram). Then, the less distant clusters are successively merged until there is just one single cluster (i.e. the root of the dendogram). Several agglomeration methods have been proposed to determine the distance between clusters, such as the maximal distance between any two observations in the clusters, the average distance between the observations in the two clusters, the distance between centroids or Ward's minimum variance criterion which minimizes the total within-cluster variance.

The unsupervised nature of these methods make them valuable for requirements selection, as they try to discover patterns into data by assigning each observation to a previously unknown group. Clustering methods can be applied based on the quantitative properties associated to requirements without human involvement. Once the requirements are clustered, that remains is to identify (or, perhaps, relate somehow) a MoSCoW category they fit in.

## 4  Experimental method design

The relationships between MoSCoW and clustering should be investigated, as a model to support the problem of requirements selection. The goal of the experimental process followed, as mentioned in the introduction, is to validate whether quantitative data about requirements (satisfaction and effort) are reliable enough to select requirements automatically by applying clustering methods (**RQ1**), including the assessment of the methods to find which gets a better solution to the problem of requirements selection (**RQ2**). Besides, a matching between groups and requirements importance categories has to be found (**RQ3**). The feasibility of tackling the requirements selection problem using clustering has been studied using both an experimental approach on three case studies and a practical one to compare the results with the manual prioritization made by some developers. The stages defined next, will be also applied to empirically assess the validity of our proposal:

i) *Prepare the data.* Clustering algorithms are affected by disparity in units,

since this influences distance computation. It can be avoided if all the variables are transformed to have a mean value of 0 and a standard deviation of 1, which makes the standard deviation the unit of measurement. As data are provided by several sources, such as software engineers (effort) or customers (values/satisfaction), different scales are used. This first standardization step ensures that experiments are performed under comparable conditions.

ii) *Estimate the number k of clusters*. Since a relation between the groups of requirements discovered by clustering methods and MoSCoW categories is expected, we set $k = 4$ (MoSCoW categories number) as the fons et origo for finding out the true relationship, cross-checking is needed (**RQ1**). The number of categories should be pore over, because other prioritization models, such as the Kano model, define different numbers of categories (**RQ3**). There are various methods for determining the optimal number of clusters $\hat{k}$, such as:

- *Elbow method* [42]. The number of clusters, $\hat{k}$, has to be chosen in a way that adding a new cluster does not produce a significant improvement in the total intra-cluster variation (or total within-cluster sum of squares), which is given by

$$WSS = \sum_{k=1}^{\hat{k}} \sum_{x_i \in C_k} (x_i - \mu_k)^2, \qquad (1)$$

where $x_i$ is an observation belonging to the cluster $C_k$ and $\mu_k$ is the mean value of the observations assigned to the cluster $C_k$. That is, the number of clusters $\hat{k}$ is chosen so that the total within-cluster sum of squares is minimized.

- *Silhouette method* [37, 27, 4]. This technique is based on the silhouette value which measures the cohesion and separation of clusters. Let $x_i$ be an observation, its silhouette $s(x_i)$ is defined as:

$$s(x_i) = \frac{a(x_i) - b(x_i)}{\max\{a(x_i), b(x_i)\}}, \qquad (2)$$

where $a(x_i)$ is the average distance between $x_i$ and the rest of observations in the same cluster, and $b(x_i)$ is the lowest average distance between $x_i$ and all observations in any other different cluster. Silhouette value, $s(x_i)$, ranges from –1 to +1, where a high value indicates that the observation is very close the ones in its cluster and very different from observations in neighbouring ones. The clustering configuration will be appropriate when most observations have a high silhouette value. However, if many observations have a low or negative value it could be either too many or too few clusters in the current clustering configuration. Thus, the average $s(x_i)$ on all observations can be considered as a measure of how appropriate the clusters are. That is, the optimal number of clusters, $\hat{k}$, should be the one that maximizes the average silhouette over a range of candidate values for $k$ [27].

- *Gap Statistic method* [43]. This technique compares the total within cluster variation with that expected under a reference null distribution of the observations, i.e. a distribution with no obvious clustering. The gap statistic for a given $k$ is defined as:

$$Gap_n(k) = E_n^* \cdot \log(W_k) - \log(W_k), \tag{3}$$

where $W_k$ is the total within-cluster sum of squares (i.e. WSS considering $k$ clusters) and $E_n^*$ denotes the expectation under a sample of size $n$ obtained from the reference distribution. The optimal number of clusters $\hat{k}$ is chosen as the smallest value of $k$ being the gap statistic within one standard deviation of the gap statistic at $k + 1$.

While elbow and silhouette methods only measure a global clustering characteristic, gap statistic formalizes the heuristics of the former methods. The optimal number of clusters $\hat{k}$ is estimated by applying the *majority rule* to the answers returned by the three methods, i.e. it is selected the value in which at least two of the three methods coincide. Once set, it will provide some insights into the answers to **RQ1** and **RQ3**.

iii) *Choose the best clustering method based on cluster validation measures.* Once the optimal number of clusters is set, the next step is to apply different clustering methods (i.e. k-means, PAM and hierarchical) and evaluate the clustering configuration found that will give an answer to **RQ2**. Ideally, we want the distance within each group to be as small as possible, keeping the average distance between groups as big as possible. Therefore, to assess a clustering configuration, internal validation measures of the clusters could be used, because they reflect the compactness, connectivity and separability of the clusters. Among the most common measures or indexes are:

- *Connectivity index* measures to what extent items are placed in the same cluster as their nearest neighbours in the observations space. For a given clustering configuration, a lower *Connectivity* indicates better clustering.
- *Dunn index* [15] aims to identify sets of clusters that are compact and well-separated. That is, when the diameter of the clusters is expected to be small and the distance between cluster to be large. For a given clustering configuration, a higher *Dunn index* indicates better clustering.
- *Silhouette index* [37] provides a validation of consistency within clusters. It is computed as the average of all observations silhouettes (see equation 2) and represents a measure of suitability with which they have been clustered. For a given clustering configuration, a higher *Silhouette index* indicates better clustering.

Table 2: Re-interpretation of MoSCoW categories in terms of the satisfaction and effort of the requirements.

|   | Category | Meaning |
|---|---|---|
| **M** | Must have | High satisfaction - Low effort |
| **S** | Should have | High satisfaction - High effort |
| **C** | Could have | Low satisfaction - Low effort |
| **W** | Won't have | Low satisfaction - High effort |

– *Caliński-Harabasz index* [11] is a variance ratio criterion to evaluate the cluster validity. It is defined as the ratio of the between-cluster variance (the variance of all cluster centroids from the centroid of the observations) to the total within-cluster variance (the average WSS of the clusters, see equation 1). Higher *Caliński-Harabasz index* values indicate better clustering, for a given clustering configuration.

Once the indicators indexes are obtained for all the clustering algorithms, the best is the one that gets the best measurement (minimizes connectivity and maximizes Dunn, Silhouette and Caliński-Harabasz indexes) (**RQ2**).

iv) *Analyse the clusters obtained and prioritize requirements.* At this point we are ready to solve the problem of requirements selection using clustering and get an answer to the research questions **RQ1**, **RQ3**. We would redefine imprecisely MoSCoW categories in terms of satisfaction and effort of the requirements, as shown in Table 2. The objective is to validate if one of the identified cluster maximizes satisfaction and minimizes effort, becoming the core of the software product as a viable product. The algorithm selected (from step iii) is run, both to find the 4 groups (suggested by MoSCoW) and the optimal number of groups (determined in step ii). Besides, a statistical summary for each group, an element representative of the members of each cluster (i.e. its centroid) and a graphical representation of the clusters, should go together to analyse the clustering configuration found. These three elements will help to get an insight into the relationship between clusters and MoSCoW categories (see Table 2), to interpret it as a mean to select requirements. Since clustering is an unsupervised technique, requirement selection will need less human effort.

## 5 Evaluation results

Three cases have been used for testing the effectiveness of our approach. The first one (20-Problem) is taken from [18]. It comprises 20 requirements and 5 customers weighed both in the 1 to 5 range. Each requirement has an associate effort scored between 1 and 10. Also, we consider implication and combination interactions between requirements. The second one (50-Problem) is taken from [2] that performs a case study of a project planned in 2008 that has been derived from the commonly known word processing tool MS Word. The raw data has been preprocessed to match problem formulation. In the case at hand, it comprises 50 requirements and 81 functional dependencies that have been elicited from 4 weighted customers. Each customer gives a value to the requirements and the effort that implies the development of each requirement has been estimated in person-hours by the development team. Finally, the third data set (100-Problem) is taken from [38]. It was generated randomly with 100 requirement, 5 customers and 44 requirements interactions, following the quantitative problem formulation given in Section 3.1. Prior to clustering, as our method indicates, the data has to be prepared. So all these data sets were standardized to avoid the influence of the disparity in the units of satisfaction and effort associated to requirements. For the sake of clarity, the next stages will be discussed regarding each data set, giving special emphasis to the cluster analysis and requirements selection, to answer the research questions.

Table 3: Evaluation of clustering methods for the 20 requirements problem

| Algorithm | K | Connectivity | Dunn | Silhouette | CH |
|---|---|---|---|---|---|
| K-means | 3 | 12.8206 | 0.2090 | 0.4666 | 22.9273 |
| PAM | 3 | 11.5250 | 0.2607 | **0.4843** | 22.6144 |
| Hierarchical | 3 | **11.1429** | 0.2576 | 0.4549 | 18.6832 |
| K-means | 4 | 20.5603 | 0.2527 | 0.4176 | **24.3832** |
| PAM | 4 | 19.9687 | **0.3151** | 0.4116 | 24.0329 |
| Hierarchical | 4 | 19.9782 | 0.2482 | 0.3561 | 18.7909 |

## 5.1 Empirical results for the 20 requirements problem

As the elbow and silhouette methods coincide (observe the vertical dashed line in Figure 2 indicating the optimal number of clusters found by each method), the majority rule dictates that the optimal number of clusters can be set at $\hat{k} = 3$. In consequence, a cluster configuration for $k = 3$ and for $k = 4$ (suggested by MoSCoW) groups are going to be explored using k-means, PAM and hierarchical clustering algorithms. Table 3 collects the values achieved by the validation indexes used and best overall values are highlighted in boldface. PAM gets the best values on two indexes (i.e. Silhouette and Dunn), which suggest its choice as the best clustering algorithm for the problem at hand.

Let us consider the MoSCoW clustering configuration (i.e. $k = 4$) found by PAM (see Figure 3.a, Table 4, and Figure 4 for a graph representation of the solution). The core of requirements identified as those that should be included in the software development project comprises $\{r_1, r_4, r_8, r_9, r_{10}, r_{11}, r_{14}, r_{15}\}$,

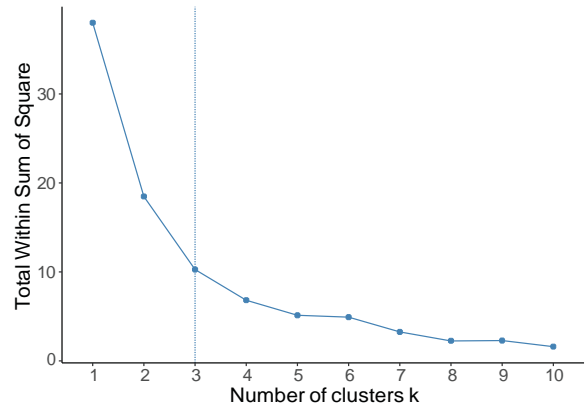

(a) Elbow method

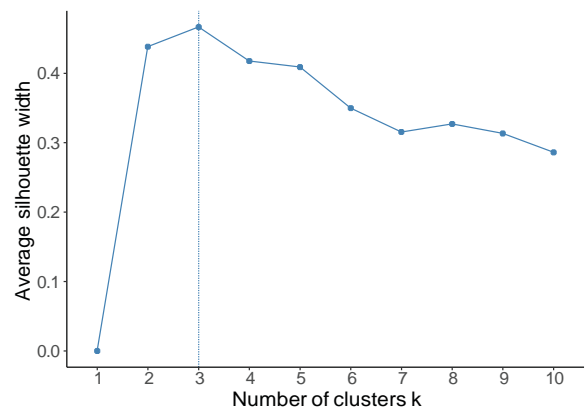

(b) Silhouette method

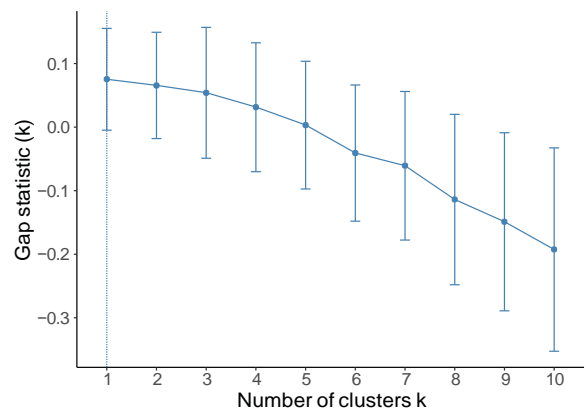

(c) Gap statistic method

Figure 2: Determining the optimal number of clusters (dashed line) for the 20-Problem

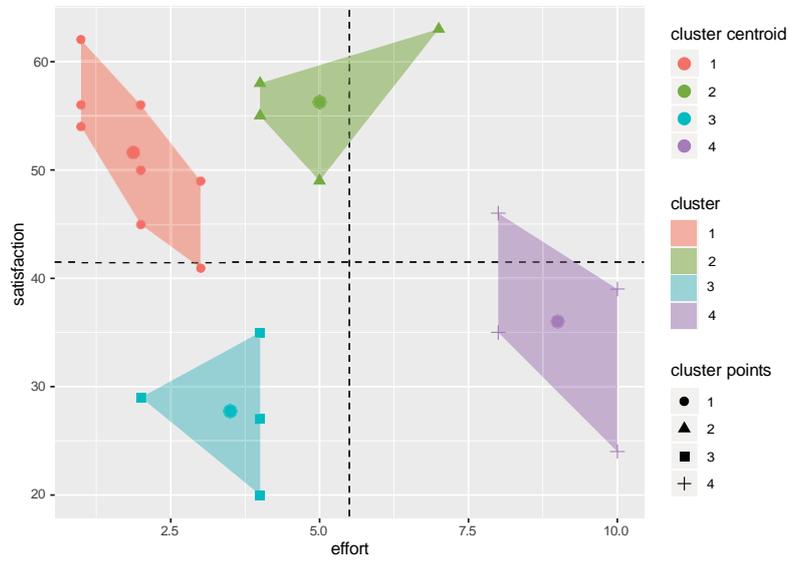

(a) Cluster configuration using the number of MoSCoW categories ($k = 4$)

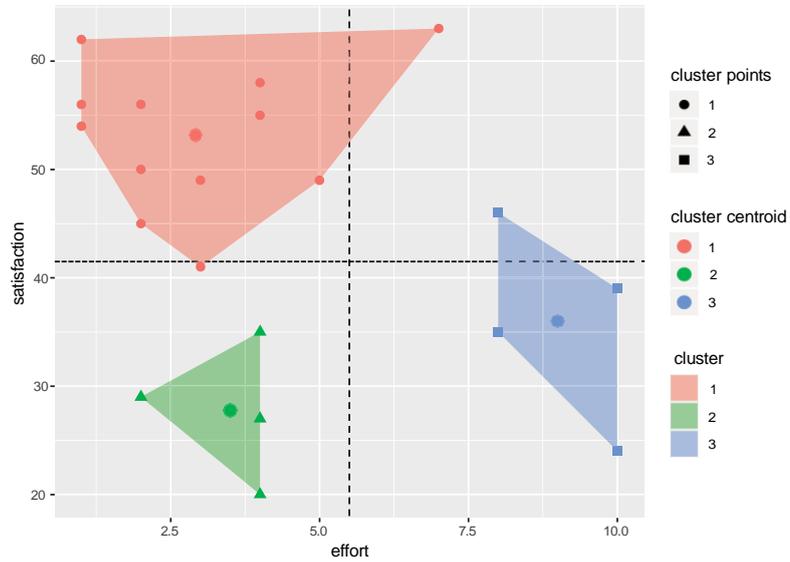

(b) Cluster configuration for the optimal number of clusters ($k = 3$)

Figure 3: Clustering for the 20-Problem

Table 4: Clusters summary (k=4) for the 20-Problem

|  | Cluster 1 | | Cluster 2 | | Cluster 3 | | Cluster 4 | |
| --- | --- | --- | --- | --- | --- | --- | --- | --- |
| Size | 8 | | 4 | | 4 | | 4 | |
|  | eff. | sat. | eff. | sat. | eff. | sat. | eff. | sat. |
| Min. | 1.00 | 41.00 | 4.0 | 49.00 | 2.0 | 20.00 | 8 | 24.00 |
| 1st. Qu. | 1.00 | 48.00 | 4.0 | 53.50 | 3.5 | 25.25 | 8 | 32.25 |
| Median | 2.00 | 52.00 | 4.5 | 56.50 | 4.0 | 28.00 | 9 | 37.00 |
| Mean | 1.87 | 51.62 | 5.0 | 56.25 | 3.5 | 27.75 | 9 | 36.00 |
| 3rd. Qu. | 2.25 | 56.00 | 5.5 | 59.25 | 4.0 | 30.50 | 10 | 40.75 |
| Max. | 3.00 | 62.00 | 7.0 | 63.00 | 4.0 | 35.00 | 10 | 46.00 |
| Centroids | 1.87 | 51.62 | 5.00 | 56.25 | 3.50 | 27.75 | 9.00 | 36.00 |

Table 5: Clusters summary (k=3) for the 20-Problem

|  | Cluster 1 | | Cluster 2 | | Cluster 3 | |
| --- | --- | --- | --- | --- | --- | --- |
| Size | 12 | | 4 | | 4 | |
|  | eff. | sat. | eff. | sat. | eff. | sat. |
| Min. | 1.000 | 41.00 | 2.0 | 20.00 | 8.0 | 24.00 |
| 1st. Qu. | 1.750 | 49.00 | 3.5 | 25.25 | 8.0 | 32.25 |
| Median | 2.500 | 54.50 | 4.0 | 28.00 | 9.0 | 37.00 |
| Mean | 2.917 | 53.17 | 3.5 | 27.75 | 9.0 | 36.00 |
| 3rd. Qu. | 4.000 | 56.50 | 4.0 | 30.50 | 10.0 | 40.75 |
| Max. | 7.000 | 63.00 | 4.0 | 35.00 | 10.0 | 46.00 |
| Centroids | 2.916 | 53.166 | 3.500 | 27.750 | 9.000 | 36.000 |

which has an associated effort value of 15 (which covers 17.65% of the total effort) and a satisfaction value of 413 (that represents 46.25% of the total satisfaction). However, if we want to get a minimum usable subset of requirements for developing a viable software product, the dependencies that exists between requirements have to be taken into account. Because of the coupling dependence $r_{11}\ \sigma_{13}$, requirement $r_{13}$ has to be included as part of the core of requirements, showed as doted node in Figure 4. Thus, the viable product obtained, showed in grey in Figure 4, has an effort value of 23 (covering 27.06% of the total effort) and a satisfaction of 448 (a 50.18% of the total satisfaction). Comparing the viable product with the core of requirements, the percentage of relative increase in satisfaction is 8.47% at the expense of a relative increase in effort of 53.33%. Figure 4 shows requirements satisfaction and effort within each node, single arrows represent implication, and double ones depict combination.

The study of the cluster configuration found by PAM using the optimal num-ber of clusters $\hat{k} = 3$ suggests (see Figure 3.b, Table 5) that the requirements core set identified is $\{r_1, r_2, r_4, r_5, r_6, r_8, r_9, r_{10}, r_{11}, r_{12}, r_{14}, r_{15}\}$ with an effort of 35 (41.18% of total effort) and satisfaction of 638 (71.45% of total satisfaction). The viable product comprises the core set enlarged with two requirements $\{r_3, r_{13}\}$ because of the coupling dependencies $r_3 \odot r_{12}$ and $r_{11} \odot r_{13}$. Thus, the viable product has an effort value of 45 (52.94% of total effort) and a satisfaction of 702 (78.61% of total satisfaction). Comparing it with the requirements core set the relative percentage increments are 10.03% in satisfaction and 28.57% in effort. It is interesting to point out that the reduction in the number of classes (from the 4 suggested by MoSCoW to the optimal number of 3) has as the result the union of the two classes in the upper left quadrant of Figure 3.a) obtained by PAM for $k = 4$, which translates into a significant increase of the viable product in terms of the number of requirements, effort and satisfaction.

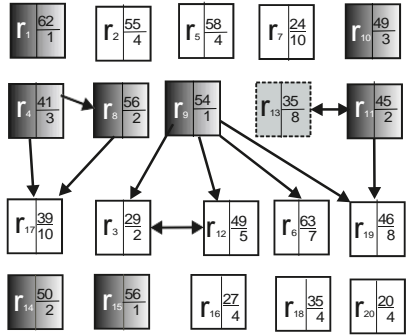

Figure 4: The viable product based on MoSCoW categories ($k = 4$). Satisfaction, effort within each node

## 5.2 Empirical results for the 50 requirements problem

When determining the optimal number of clusters, all methods (elbow, silhouette and gap statistic) coincide (see Figure 5) and indicate that $k = 4$ should be the number of clusters. Then, only the MoSCoW cluster configuration will be found and analysed. The validation indexes values achieved by the different clustering algorithms are shown in Table 6 (the best overall values are highlighted in boldface). *K-means* is the best clustering algorithm, whose cluster configuration is shown in Figure 6, see also Table 7 for detailed data.

The identified requirements core set is $\{r_3, r_{12}, r_{13}, r_{14}, r_{15}, r_{16}, r_{17}, r_{20}, r_{21}, r_{22}, r_{24}, r_{27}, r_{28}, r_{29}, r_{30}, r_{31}, r_{32}, r_{33}, r_{34}, r_{35}, r_{39}, r_{44}, r_{45}, r_{46}, r_{47}, r_{49}, r_{50}\}$, which has to be enlarged with $\{r_1, r_2, r_4, r_5, r_6, r_7, r_8, r_9, r_{10}, r_{11}, r_{19}, r_{38}, r_{43}\}$, because of implication dependencies to resemble a viable product that is shown in Figure 7, where single arrows to represent implication, and double ones for combina-tion. The requirements core set has an effort value of 122 (26.41% of total effort) and a satisfaction of 3913 (60.41% of total satisfaction). Once dependencies are taken into account, the viable product has an associated effort of 314 (67.97% of total effort) and satisfaction of 5700 (87.99% of total satisfaction), which, with respect to the core set, supposes an increase of 45.67% in satisfaction at the expense of increasing the effort by 157.38%. As it can be seen in Table 8, the set of functionalities that has not being included in the project comprises complementary views for the document, external links and specific location editing options, text formatting (change case and document background), additional tools and data import.

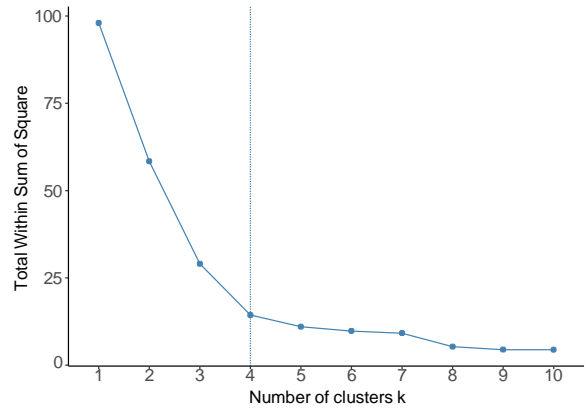

(a) Elbow method

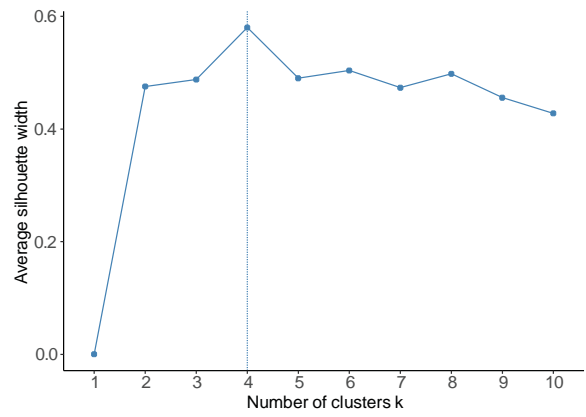

(b) Silhouette method

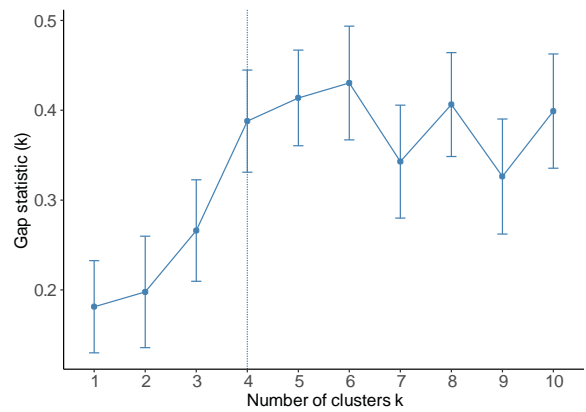

(c) Gap statistic method

Figure 5: Determining the optimal number of cluster for the 50-Problem

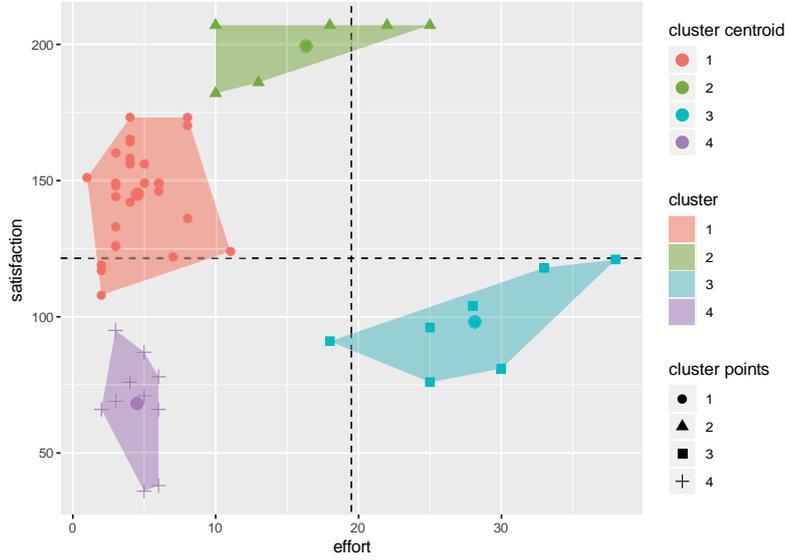

Figure 6: Clustering for the 50-Problem

Table 6: Evaluation of clustering methods for the 50-Problem

| Algorithm | K | Connectivity | Dunn | Silhouette | CH |
|---|---|---|---|---|---|
| K-means | 4 | **14.3071** | **0.1317** | **0.5801** | **89.1628** |
| PAM | 4 | 15.7456 | 0.0909 | 0.5765 | 87.6092 |
| Hierarchical | 4 | 16.0647 | 0.1249 | 0.5655 | 83.4536 |

### 5.3 Empirical results for the 100 requirements problem

Figure 8 shows that the elbow and silhouette methods coincide in the optimal number of clusters $\hat{k}$ and, it is set to 3. As it differs from the number of MoSCoW categories ($k = 4$), the clusters configurations using both values of $k$ will be explored. According to the validation indexes (see Table 9), *hierarchical* clustering gets two, out of four, best values for the indexes, becoming the selected algorithm.

Table 7: Clusters summary (k=4) for the 50-Problem

|  | Cluster 1 | | Cluster 2 | | Cluster 3 | | Cluster 4 | |
| --- | --- | --- | --- | --- | --- | --- | --- | --- |
| Size | 27 | | 6 | | 7 | | 10 | |
|  | eff. | sat. | eff. | sat. | eff. | sat. | eff. | sat. |
| Min. | 1.00 | 108.0 | 10.00 | 182.0 | 18.00 | 76.00 | 2.00 | 36.0 |
| 1st. Qu. | 3.00 | 129.5 | 10.75 | 191.2 | 25.00 | 86.00 | 3.25 | 66.0 |
| Median | 4.00 | 149.0 | 15.50 | 207.0 | 28.00 | 96.00 | 5.00 | 70.0 |
| Mean | 4.52 | 144.9 | 16.33 | 199.3 | 28.14 | 98.14 | 4.50 | 68.2 |
| 3rd. Qu. | 6.00 | 157.0 | 21.00 | 207.0 | 31.50 | 111.00 | 5.75 | 77.5 |
| Max. | 11.00 | 173.0 | 25.00 | 207.0 | 38.00 | 121.00 | 6.00 | 95.0 |
| Centroids | 4.52 | 144.92 | 16.33 | 199.33 | 28.14 | 98.14 | 4.50 | 68.20 |

The MoSCoW clustering configuration found by hierarchical clustering (see Figure 9.a, Table 10), identified cluster 4, $\{r_8, r_9, r_{15}, r_{16}, r_{18}, r_{19}, r_{20}, r_{21}, r_{23}, r_{24}, r_{25}, r_{27}, r_{29}, r_{32}, r_{34}, r_{37}, r_{40}, r_{41}, r_{46}, r_{49}, r_{50}, r_{51}, r_{52}, r_{54}, r_{56}, r_{60}, r_{62}, r_{68}, r_{72}, r_{73}, r_{77}, r_{89}, r_{90}, r_{93}, r_{94}, r_{98}, r_{100}\}$ as the requirements core set. This set has an associated effort value of 218 (21.02% of total effort) and a satisfaction value of 1089 (40% of total satisfaction). From it the viable product is completed by adding the requirements set $\{r_2, r_3, r_{10}, r_{14}, r_{22}, r_{30}, r_{33}, r_{47}\}$, since the problem dependencies have to be fulfilled. Its effort and satisfaction values are 345 (33.27% of total effort) and 1282 (48.27% of total satisfaction), respectively. The relative percentage increments with respect to the requirements core set are 58.26% in effort and 17.72% in satisfaction.

However, if we use hierarchical clustering with the optimal number of clusters $\hat{k} = 3$ (see Figure 9.b, Table 11), we get as requirements core set the union of clusters 4 and 3 found previously for $k = 4$ (Table 10 shows data for $\hat{k} = 4$). Once dependencies are taken into account, the viable product has an associated effort of 477 (50% of total effort) and satisfaction of 1733 (65.25% of total satisfaction). With respect to the core set the relative percentage increment in effort is 44.99% and 14.01% in satisfaction. Therefore, the viable product will enlarge that found for MoSCoW ($k = 4$) and will be larger in terms of the number of requirements, effort and satisfaction.

## 5.4 Comparison with developers' classification

An additional validation has been made comparing our proposal with the subjective developers' prioritization. The 50-Problem is our testing scenario be-

Table 8: 50-Problem - Requirements Description

| ID | | Short name | Description |
|---|---|---|---|
| 1 | File-1 (*) | New File | Crete a New File |
| 2 | File-2 (*) | Open File | Open an Existing File |
| 3 | **File-3** | **Close File** | Close Current File |
| 4 | File-4 (*) | Save File | Save a File |
| 5 | File-5 (*) | Save as | Save a File as a Different File Type |
| 6 | File-6 (*) | Search File | Search for a File in The computer containing some Text |
| 7 | File-7 (*) | Protect File | Make a File Password Protected |
| 8 | File-8 (*) | Print Preview | Print Preview a File |
| 9 | File-9 (*) | Print File | Print Current File |
| 10 | File-10 (*) | Send To | Send File To Email/FAX |
| 11 | File-11 (*) | Set Properties | Set File Header Inform ton |
| 12 | **File-12** | **Exit** | Save and Exit from Word Processing Application |
| 13 | **Edit-1** | **Undo a Task** | Undo a Task and goes back to previous state |
| 14 | **Edit-2** | **Redo a Task** | Redo the most recent Change |
| 15 | **Edit-3** | **Cut** | Delete a Text and Copy to Clipboard |
| 16 | **Edit-4** | **Copy** | Copy a Text |
| 17 | **Edit-5** | **Paste** | Paste a Text from Clipboard |
| 18 | Edit-6 | Paste Special | External Linking and Embedding |
| 19 | Edit-7 (*) | Go To | Go to a specific location in the current file |
| 20 | **Edit-8** | **Find** | Search for a Text in the current Document |
| 21 | **Edit-9** | **Replace** | Search and Replace a Text |
| 22 | **Edit-10** | **Select All** | Select All From Current File |
| 23 | View-1 | Default | Switch to Default View |
| 24 | **View-2** | **Print Layout** | Print Layout View of Current File |
| 25 | View-3 | Web layout | Web Page Layout View of Current File |
| 26 | View-4 | Zoom | Zoom In/Out |
| 27 | **View-5** | **Header/Footer** | Show Healer/Footer of Current File |
| 28 | **Insert-1** | **Page Numbers** | Insert Page Numbers in Footer |
| 29 | **Insert-2** | **Date/Time** | Insert Date/Time in the File Footer |
| 30 | **Insert-3** | **Symbol** | Insert Symbol to the Cursor Location |
| 31 | **Insert-4** | **Bookmark** | Set/Update File Bookmark |
| 32 | **Insert-5** | **Hyperlink** | Insert a Hyperlinked Text |
| 33 | **Format-1** | **Font** | Change Font Setting of Selected Text |
| 34 | **Format-2** | **Paragraph** | Set the Paragraph Formatting |
| 35 | **Format-3** | **Bullets/Numbers** | Insert Bullets/Numbering to Selected Text |
| 36 | Format-4 | Change Case | Change to Upper/Lower/Mixed Case selected Text |
| 37 | Format-5 | Background | Change Background of the Document |
| 38 | Tools-1 (*) | Ch-Spell | Spelling Check Tool |
| 39 | **Tools-2** | **Ch-Grammar** | Grammar Check Tool |
| 40 | Tools-3 | Speech | Read the Text of The Document |
| 41 | Tools-4 | Mail Merge | Mail Merge with Existing Customer Database |
| 42 | Tools-5 | Macro | Create/Use Automated Document Processing Function |
| 43 | Tools-6 (*) | Set Options | Configure Documents Options |
| 44 | **Data-1** | **Insert Table** | Insert a Table |
| 45 | **Data-2** | **Delete Table** | Delete an Existing Table |
| 46 | **Data-3** | **Table Format** | Format a Existing Table |
| 47 | **Data-4** | **Sort** | Sort Selected Data |
| 48 | Data-5 | Import Data | Import Data from External Database |
| 49 | **Help-1** | **Help** | Loads Applicant Help File |
| 50 | **Help-2** | **Search** | Searches a Text in the Document Help File |

core requirements in bold

∗ requirements included by effect of requirements implication and coupling

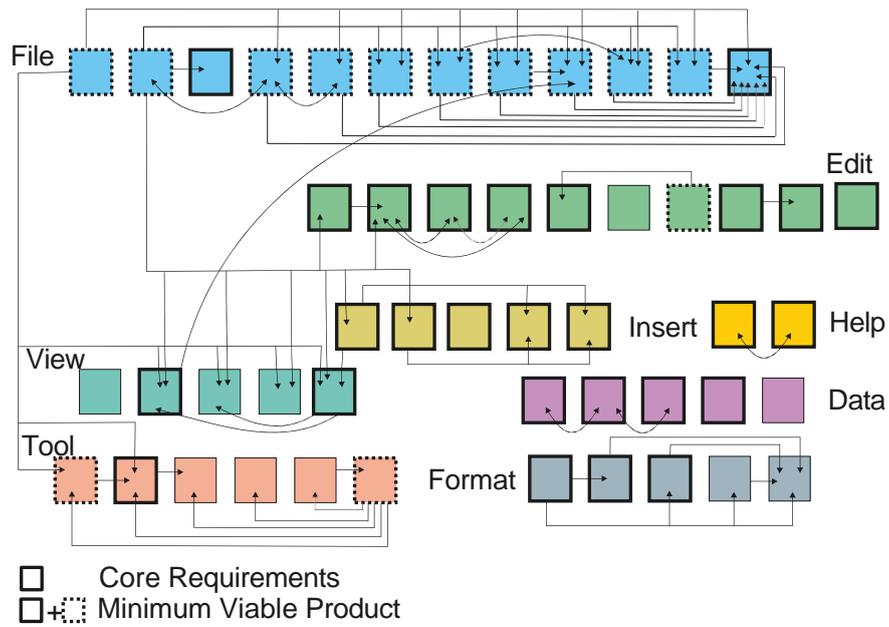

Figure 7: The viable product for 50-Problem

Table 9: Evaluation of clustering methods for the 100-Problem

| Algorithm | K | Connectivity | Dunn | Silhouette | CH |
|---|---|---|---|---|---|
| K-means | 3 | 17.7056 | 0.0548 | 0.4283 | 89.5132 |
| PAM | 3 | 13.4373 | 0.0831 | **0.4308** | 89.3966 |
| Hierarchical | 3 | **7.2357** | **0.1096** | 0.4278 | 88.0933 |
| K-means | 4 | 18.9746 | 0.0783 | 0.3993 | **90.9959** |
| PAM | 4 | 26.2714 | 0.0696 | 0.3993 | 88.7641 |
| Hierarchical | 4 | 14.8242 | 0.1096 | 0.3964 | 82.5902 |

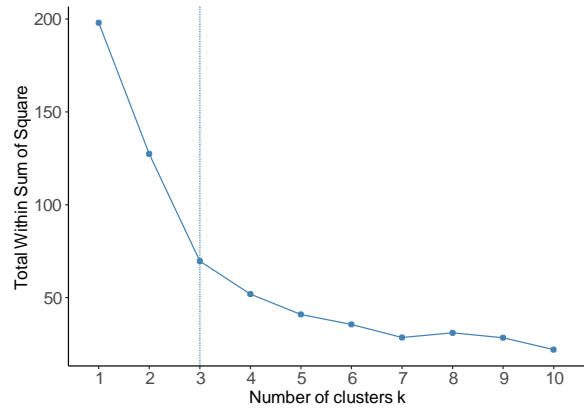

(a) Elbow method

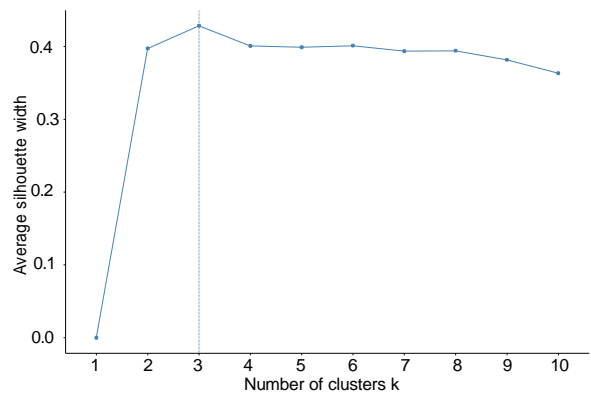

(b) Silhouette method

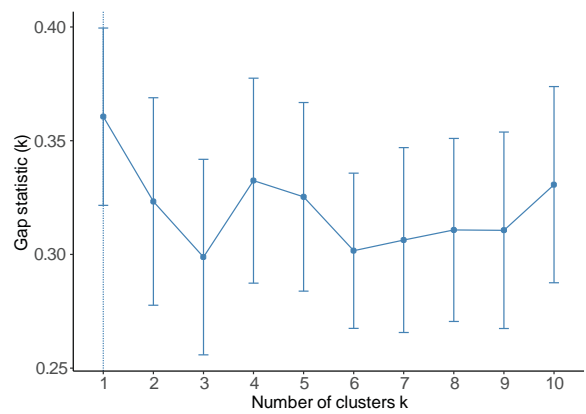

(c) Gap statistic method

Figure 8: Determining the optimal number of cluster for the 100-Problem

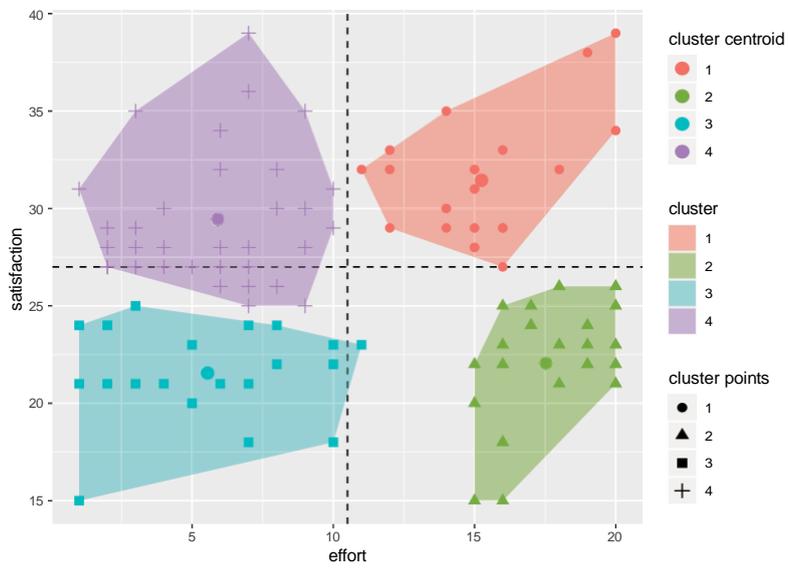

(a) Cluster configuration using the number of MoSCoW categories ($k = 4$)

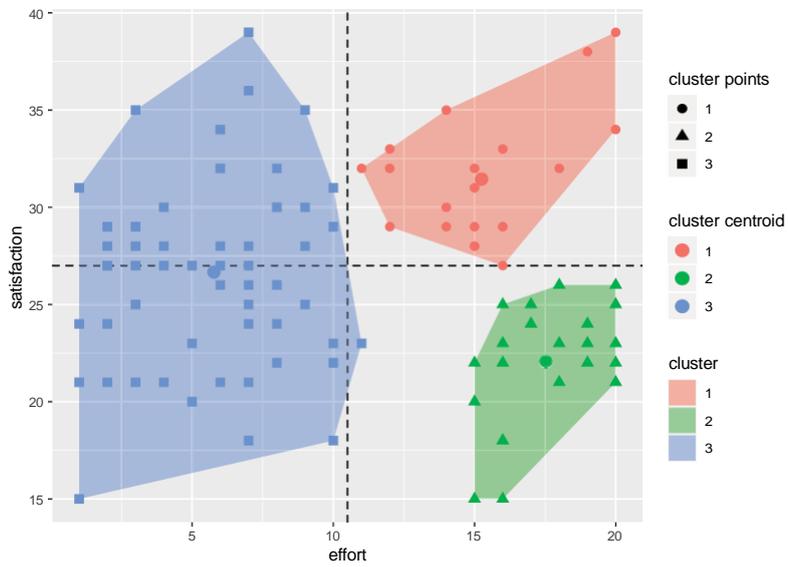

(b) Cluster configuration for the optimal number of clusters ($k = 3$)

Figure 9: Clustering for the 100-Problem

Table 10: Clusters summary (k=4) for the 100-Problem

|  | Cluster 1 | | Cluster 2 | | Cluster 3 | | Cluster 4 | |
| --- | --- | --- | --- | --- | --- | --- | --- | --- |
| Size | 20 | | 23 | | 20 | | 37 | |
|  | eff. | sat. | eff. | sat. | eff. | sat. | eff. | sat. |
| Min. | 11.00 | 27.00 | 15.00 | 15.00 | 1.00 | 15.00 | 1.000 | 25.00 |
| 1st. Qu. | 14.00 | 29.00 | 16.00 | 21.00 | 2.75 | 21.00 | 4.000 | 27.00 |
| Median | 15.00 | 31.50 | 17.00 | 23.00 | 5.50 | 21.50 | 6.000 | 28.00 |
| Mean | 15.25 | 31.45 | 17.52 | 22.04 | 5.55 | 21.55 | 5.892 | 29.43 |
| 3rd. Qu. | 16.00 | 33.00 | 19.00 | 24.00 | 8.00 | 23.25 | 8.000 | 31.00 |
| Max. | 20.00 | 39.00 | 20.00 | 26.00 | 11.00 | 25.00 | 10.000 | 39.00 |
| Centroids | 15.25 | 31.45 | 17.52 | 22.04 | 5.55 | 21.55 | 5.89 | 29.43 |

Table 11: Clusters summary (k=3) for the 100-Problem

|  | Cluster 1 | | Cluster 2 | | Cluster 3 | |
| --- | --- | --- | --- | --- | --- | --- |
| Size | 20 | | 23 | | 57 | |
|  | eff. | sat. | eff. | sat. | eff. | sat. |
| Min. | 11.00 | 27.00 | 15.00 | 15.00 | 1.000 | 15.00 |
| 1st. Qu. | 14.00 | 29.00 | 16.00 | 21.00 | 3.000 | 23.00 |
| Median | 15.00 | 31.50 | 17.00 | 23.00 | 6.000 | 27.00 |
| Mean | 15.25 | 31.45 | 17.52 | 22.04 | 5.772 | 26.67 |
| 3rd. Qu. | 16.00 | 33.00 | 19.00 | 24.00 | 8.000 | 30.00 |
| Max. | 20.00 | 39.00 | 20.00 | 26.00 | 11.000 | 39.00 |
| Centroids | 15.250 | 31.450 | 17.521 | 22.043 | 5.771 | 26.666 |

cause it is the only data set that includes requirements descriptions. This allows us to compare the automatic prioritization with a manual one made by developers. This comparison has certain risks such as the unknown skill level of researchers/participants or the different perceptions of the problem [29]. Nonetheless, this alternative validation approach could clarify if clustering is efficacious enough to classify requirement using MoSCoW model. Seven senior software engineering degree students, having no information on requirements dependencies, were requested to apply MoSCoW and get a prioritization of requirements. They relied only on requirements descriptions to come up with a prioritization without offering any value based rationale for making the decision about how to rate the priority [46].

Afterwards, the *majority rule* (i.e. the MoSCoW category selected was the one indicated more times by the students) was used to get the final students' classification shown on Figure 10 and the tie-breaker is the order of the MoSCoW classes. The first finding worth highlighting is that there isn't a clear agreement between developers on the requirements that should be included in the 'must have' set. Requirements marked with '*', are those whose class is agreed by half or more of the developers.

The requirements core set agreed by developers is $\{r_1, r_2, r_3, r_4, r_5 r_9, r_{12}, r_{13}, r_{14}, r_{15}, r_{16}, r_{17}, r_{20}, r_{26}, r_{28}, r_{31}, r_{33}, r_{38}, r_{39}, r_{44}, r_{45}, r_{46}, r_{49}, r_{50}\}$. From it we get the viable product by adding the requirements set $\{r_6, r_7, r_8, r_{10}, r_{11}, r_{19}, r_{24}, r_{29}, r_{43}\}$ since the dependencies of the problem have to be fulfilled. Requirements with a '1' in the bottom of Figure 10 are those included in the core set, whereas those added due to dependencies are marked with a '2'. The core set has an value of 190 (41.13% of total effort) and a satisfaction of 3654 (56.41% of total satisfaction). After dependencies have been taken into account to obtain a viable product, the effort and satisfaction values change to 285 (61.69% of total effort) and 4626 (71.41% of total satisfaction), respectively. The relative percentage change with respect to the requirements core set is 50% for effort and 26.6% for satisfaction. It can be observed that developers include fewer requirements than clustering. An explanation for this could be that their prioritization is made based on natural language descriptions instead of effort and satisfaction values. Although in both approaches dependencies are needed to get the viable product from the core set of requirements, that got by developers has lower effort and satisfaction values than in the case of clustering. This fact suggests that clustering (Must have cluster, see table 2) could be a good starting point for requirements negotiation (see Fig 1).

Experimental results obtained on the three data sets reinforce the idea that requirements effort and satisfaction can be used as measures for clustering requirements and to define the group of requirements that must be included into a successful next software release. That is to say, automatic selection of requirements can be approached applying clustering techniques (**RQ1**). The decision making processes related to software negotiation and conflict resolution can be successfully assisted by clustering the scored requirements attributes, see Figure 1. However, there is no single applicable clustering algorithm. Which is the one to be applied depends on the problem (**RQ2**). Although the optimal number of clusters $\hat{k}$ differs between the case studies, this discrepancy can be interpreted, from the requirements' selection point of view, as the union of the two higher priority requirements groups (table 2 (**RQ3**), which is also advocated by other models as Kano. What is worth to highlight here is that, independently of the case at hand, clustering methods can identify a core set of requirements. However, requirements dependencies have to be managed at the time of getting a viable product, when additional requirements have to be incorporated to the core to fulfill dependencies. The comparison with manual prioritization suggests that the subjective nature of the problem makes difficult to reach an agreement, making interesting the use or any automatic approach, such as clustering.

# 6  Limitations and threats to validity

The validity of any research work indicates how trustworthy and generalizable their results are [47]. The potential threats are discussed next, describing also what has been done to mitigate them.

In this paper, since the objects studied are requirements and their associated quantitative data (satisfaction and effort) are estimated and provided by software engineers and customers, a construct validity threat could appear because of the subjectivity and accuracy of the estimations. Although development teams use estimations in everyday work, to assure that experiments are performed under a common unified framework, we prepare data as the first step before clustering, as it has been described in section 4, so as to unify value ranges.

A potential threat to internal validity is related to the experimental methodology that has been applied. Since there are different clustering approaches, a threat that arises is which one to use. The methodology applied mitigates this by considering several representative clustering approaches and choosing the best based on cluster validation measures, as they reflect the compactness, connectivity and separability of the clusters found. Related to validation measures several issues come up such as what measures to use and, based on those, how to choose the algorithm and the optimal number of groups. To deal with these issues, we resort to the *majority rule* on validation measures selected in a way that they cover the three characteristics to be studied in the clusters found.

Results (i.e. conclusion validity threat) have been interpreted by taking into consideration, besides the aspects related to clustering, requirements dependencies as the primary goal is to select requirements to get a core set of requirements for the software product. This is also related to including more important functionalities and can affect to the number of requirements groups identified. Finally, regarding external validity, we use different data sets to analyse the generalization of the observed results.

# 7 Conclusions

The core set of requirements is a software increment with just enough features to satisfy customers. The definition of which requirements should be included in it is a critical decision process in software product development that most times entails striking a balance between the value added to the project outcome and their cost. Requirements selection and negotiation has to be conducted according to these values and requirements dependencies. Clustering methods can provide a valuable aid by finding a requirements group that will make up the core of the next software product.

We have proposed a methodology for requirements selection based on clustering techniques. Through its use, we can define the requirements core set that has to be incorporated into a next software release. From this set, we have obtained a viable product by taking into account requirements dependencies and incorporating to the core set those requirements needed to fulfil them. This will be the starting requirements pack used in the contractual negotiations with stakeholders at the time of defining the project scope. The feasibility of our methodology has been tested on several case studies, following the established guidelines that define an experimentation protocol in other cases when the suitability of clustering strategies needs to be studied.

We have learned several lessons. No single clustering algorithm applies, as it can be seen in the cases studied when a specific clustering method has to be selected. In each case a different clustering algorithm were selected. The number of categories proposed by MoSCoW can serve as a good starting point for requirements selection. In the cases studied, the effect of the optimal number of clusters has been to enlarge both the requirements core set and the viable product, increasing satisfaction and effort. Dependencies have to be taken into account always when identifying a viable product blurring clusters boundaries. When comparing manual and automatic prioritization, results suggest that developers include fewer requirements which translates into a viable product that needs less development effort but with less satisfaction.

# acknowledgements

This research has been financed by the Spanish Ministry of Science, Innovation and Universities under the Search based software engineering research network (RED2018-102472-T), by the Spanish Ministry of Economy and Competitiveness under project TIN2016-77902-C3-3-P (PGM-SDA II project) and it is partially supported by Data, Knowledge and Software Engineering (DKSE) research group (TIC-181) of the University of Almer´ıa.

# References

[1] Achimugu, P., Selamat, A., Ibrahim, R., Mahrin, M.N.: A systematic literature review of software requirements prioritization research. Information and software technology **56**(6), 568–585 (2014). DOI https://doi.org/10.1016/j.infsof.2014.02.001

[2] Agarwal, N., KariFmpour, R., Ruhe, G.: Theme-based product release planning: An analytical approach. In: 2014 47th Hawaii International Conference on System Sciences, pp. 4739–4748. IEEE (2014). DOI https://doi.org/10.1109/HICSS.2014.582

[3] del Águila, I.M., del Sagrado, J.: Three steps multiobjective decision process for software release planning. Complexity **21**(S1), 250–262 (2016). DOI https://doi.org/10.1002/cplx.21739

[4] Amorim, R.C.d., Hennig, C.: Recovering the number of clusters in data sets with noise features using feature rescaling factors. Information Sciences **324**, 126–145 (2015). DOI 10.1016/j.ins.2015.06.039

[5] Bagnall, A.J., Rayward-Smith, V.J., Whittley, I.: The next release problem. Information & Software Technology **43**(14), 883–890 (2001). DOI https://doi.org/10.1016/S0950-5849(01)00194-X

[6] Beck, K.: Extreme programming explained: embrace change. Addison-Wesley Professional (2000)

[7] Berander, P., Andrews, A.: Requirements Prioritization, pp. 69–94. Springer Berlin Heidelberg, Berlin, Heidelberg (2005). DOI https://doi.org/10.1007/3-540-28244-0 4

[8] Bourque, P., Fairley, R.E. (eds.): SWEBOK: Guide to the Software Engineering Body of Knowledge, version 3.0 edn. IEEE Computer Society, Los Alamitos, CA (2014)

[9] Brennan, K. (ed.): A Guide to the Business Analysis Body of Knowledge (BABOK Guide), Version 2.0. International Institute of Business Analysis (2009)

[10] Bühne, S., Herrmann, A.: Handbook Requirements Management according to the IREB Standard. International Requirements Engineering Board, IREB, 1 edn. (2019). Education and Training for the IREB Certified Professional for Requirements Engineering Qualification Advanced Level Requirements Management

[11] Caliński, T., Harabasz, J.: A dendrite method for cluster analysis. Communications in Statistics **3**(1), 1–27 (1974). DOI 10.1080/03610927408827101

[12] Carlshamre, P., Sandahl, K., Lindvall, M., Regnell, B., et al.: An industrial survey of requirements interdependencies in software product release planning. In: Requirements Engineering, 2001. Proceedings. Fifth IEEE International Symposium on, pp. 84–91. IEEE (2001). DOI https://doi.org/10.1109/ISRE.2001.948547

[13] Davis, A.M.: The art of requirements triage. Computer **36**(3), 42–49 (2003). DOI https://doi.org/10.1109/MC.2003.1185216

[14] Domínguez-Ríos, M.A., Chicano, F., Alba, E., del Águila, I., del Sagrado, J.: Efficient anytime algorithms to solve the bi-objective next release problem. Journal of Systems and Software **156**, 217 – 231 (2019). DOI https://doi.org/10.1016/j.jss.2019.06.097

[15] Dunn, J.: A fuzzy relative of the isodata process and its use in detecting compact well-separated clusters. Journal of Cybernetics **3**(3), 32–57 (1973). DOI 10.1080/01969727308546046

[16] El Emam, K., Koru, A.G.: A replicated survey of it software project failures. IEEE software **25**(5), 84–90 (2008)

[17] Franch, X., Ruhe, G.: Software release planning. In: Proceedings of the 38th International Conference on Software Engineering Companion, pp. 894–895 (2016). DOI 10.1145/2889160.2891051

[18] Greer, D., Ruhe, G.: Software release planning: an evolutionary and iterative approach. Information and software technology **46**(4), 243–253 (2004). DOI https://doi.org/10.1016/j.infsof.2003.07.002

[19] Group, S.: The chaos report. Tech. rep., Standish Group (1995)

[20] Harman, M., Krinke, J., Medina-Bulo, I., Palomo-Lozano, F., Ren, J., Yoo, S.: Exact scalable sensitivity analysis for the next release problem. ACM Transactions on Software Engineering and Methodology (TOSEM) **23**(2), 19 (2014). DOI https://doi.org/10.1145/2537853

[21] Harman, M., Mansouri, S.A., Zhang, Y.: Search-based software engineering: Trends, techniques and applications. ACM Comput. Surv. **45**(1), 11:1–11:61 (2012). DOI https://doi.org/10.1145/2379776.2379787

[22] Hartigan, J., Wong, M.: Algorithm as 136: A k-means clustering algorithm. Journal of the Royal Statistical Society, Series C (Applied Statistics) **28**(1), 100 – 108 (1979). DOI https://doi.org/10.2307/2346830

[23] Hujainah, F., Bakar, R.B.A., Abdulgabber, M.A., Zamli, K.Z.: Software requirements prioritisation: A systematic literature review on significance, stakeholders, techniques and challenges. IEEE Access (2018)

[24] Jung, H.W.: Optimizing value and cost in requirements analysis. IEEE Software **15**(4), 74–78 (1998). DOI https://doi.org/10.1109/52.687950

[25] Kano, N.: Attractive quality and must-be quality. Hinshitsu (Quality, the Journal of Japanese Society for Quality Control) **14**, 39–48 (1984)

[26] Karlsson, J., Olsson, S., Ryan, K.: Improved practical support for large-scale requirements prioritising. Requirements Engineering **2**(1), 51–60 (1997). DOI https://doi.org/10.1007/BF02802897

[27] Kaufman, L., Rousseeuw, P.J.: Finding Groups in Data: An Introduction to Cluster Analysis. Wiley, New York (1990)

[28] Leffingwell, D.: Managing software requirements: a use case approach. Pearson Education India (2003)

[29] Lethbridge, T.C., Lyon, S., Perry, P.: The management of university–industry collaborations involving empirical studies of software engine. In: Guide to Advanced Empirical Software Engineering, pp. 257–281. Springer (2008)

[30] Maalej, W., Nayebi, M., Johann, T., Ruhe, G.: Toward data-driven requirements engineering. IEEE Software **33**(1), 48–54 (2016). DOI https://doi.org/10.1109/MS.2015.153

[31] MacQueen, J.: Some methods for classification and analysis of multivariate observations. In: L.M.L. Cam, J. Neyman (eds.) Proceedings of the Fifth Berkeley Symposium on Mathematical Statistics and Probability, pp. 281–297. Berkeley, CA: University of California Press (1967)

[32] Matzler, K., Hinterhuber, H.H.: How to make product development projects more successful by integrating kano's model of customer satisfaction into quality function deployment. Technovation **18**(1), 25 – 38 (1998). DOI https://doi.org/10.1016/S0166-4972(97)00072-2

[33] Murtagh, F., Contreras, P.: Algorithms for hierarchical clustering: an overview. WIREs Data Mining and Knowledge Discovery **2**(1), 86–97 (2012). DOI https://doi.org/10.1002/widm.53

[34] Nayebi, M., Ruhe, G.: Asymmetric release planning - compromising satisfaction against dissatisfaction. IEEE Transactions on Software Engineering pp. 1–1 (2018). DOI 10.1109/TSE.2018.2810895

[35] Ngo-The, A., Ruhe, G., Shen, W.: Release planning under fuzzy effort constraints. In: Cognitive Informatics, 2004. Proceedings of the Third IEEE International Conference on, pp. 168–175. IEEE (2004). DOI https://doi.org/10.1109/COGINF.2004.1327472

[36] Pohl, K., Rupp, C.: Requirements engineering fundamentals: a study guide for the certified professional for requirements engineering exam-foundation level-IREB compliant. Rocky Nook, Inc. (2016)

[37] Rousseeuw, P.J.: Silhouettes: A graphical aid to the interpretation and validation of cluster analysis. Journal of Computational and Applied Mathematics **20**, 53–65 (1987). DOI 10.1016/0377-0427(87)90125-7

[38] del Sagrado, J., del Águila, I.M., Orellana, F.J.: Multi-objective ant colony optimization for requirements selection. Empirical Software Engineering **20**(3), 577–610 (2015). DOI https://doi.org/10.1007/s10664-013-9287-3

[39] Sauerwein, E., Bailom, F., Matzler, K., Hinterhuber, H.H.: The kano model: How to delight your customers. In: International Working Seminar on Production Economics, vol. 1, pp. 313–327 (1996)

[40] Svahnberg, M., Gorschek, T., Feldt, R., Torkar, R., Saleem, S.B., Shafique, M.U.: A systematic review on strategic release planning models. Information and software technology **52**(3), 237–248 (2010). DOI https://doi.org/10.1016/j.infsof.2009.11.006

[41] Thakurta, R.: Understanding requirement prioritization artifacts: a systematic mapping study. Requirements engineering **22**(4), 491–526 (2017). DOI https://doi.org/10.1007/s00766-016-0253-7

[42] Thorndike, R.: Who belongs in the family? Psychometrika **18**, 267–276 (1953). DOI https://doi.org/10.1007/BF02289263

[43] Tibshirani, R., Walther, G., Hastie, T.: Estimating the number of clusters in a data set via the gap statistic. Journal of the Royal Statistical Society B **63**(2), 411–423 (2001). DOI 10.1111/1467-9868.00293

[44] Van den Akker, J.M., Brinkkemper, S., Diepen, G., Versendaal, J.: Determination of the next release of a software product: an approach using integer linear programming. In: Proceeding of the 11th International Workshop on Requirements Engineering: Foundation for Software Quality REFSQ'05, pp. 247–262 (2005)

[45] Wagner, S., Fernández, D.M., Felderer, M., Vetrò, A., Kalinowski, M., Wieringa, R., Pfahl, D., Conte, T., Christiansson, M.T., Greer, D., et al.: Status quo in requirements engineering: A theory and a global family of surveys. ACM Transactions on Software Engineering and Methodology (TOSEM) **28**(2), 9 (2019)

[46] Wiegers, K., Beatty, J.: Software Requirements, 3e. Microsoft Press (2013)

[47] Wohlin, C., Runeson, P., Höst, M., Ohlsson, M.C., Regnell, B., Wesslén, A.: Experimentation in software engineering. Springer Science & Business Media (2012)

[48] Zhang, Y., Harman, M., Ochoa, G., Ruhe, G., Brinkkemper, S.: An empirical study of meta-and hyper-heuristic search for multi-objective release planning. ACM Transactions on Software Engineering and Methodology (TOSEM) **27**(1), 3 (2018)

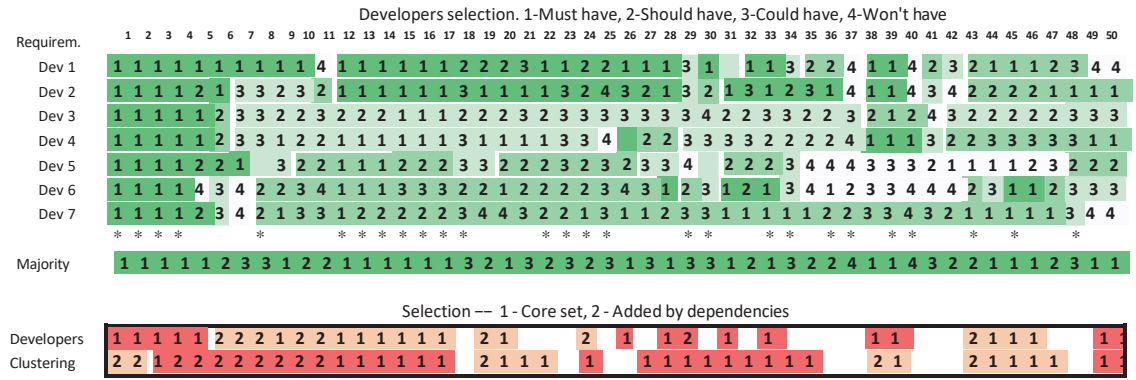

Figure 10: Developers selection for 50-Problem